\documentclass[5p]{elsarticle}

\usepackage{graphicx}
\usepackage{amsmath}
\usepackage{siunitx}
\usepackage[utf8]{luainputenc}
\usepackage{todonotes}

\usepackage{braket}
\usepackage[hidelinks]{hyperref}

\newcommand{\arccot}{\text{arccot}}

\newcommand{\expect}[1]
{
\langle #1 \rangle
}

\begin{document}

\begin{frontmatter} 
\title{ Entanglement and Entropy in Electron-Electron Scattering
}
\author[ifp,ustem]{P. Schattschneider\corref{cor1}}
\ead{peter.schattschneider@tuwien.ac.at}
\cortext[cor1]{Corresponding author}
\author[ustem]{S. Löffler}
\author[udu]{H. Gollisch}
\author[udu]{R. Feder}

\address[ifp]{Institute of Solid State Physics, TU Wien, Wiedner Hauptstraße 8-10/E138, 1040 Wien, Austria}
\address[ustem]{University Service Centre for Transmission Electron Microscopy, TU Wien, Wiedner Hauptstraße 8-10/E138, 1040 Wien, Austria}
\address[udu]{Universit\"at Duisburg-Essen, D-47057 Duisburg, Germany} 

\begin{abstract}
Treating Coulomb scattering of two free electrons in a stationary approach,
we explore the momentum and spin entanglement created by the 
interaction.
We show that a particular discretisation  provides an 
estimate of the von Neumann entropy of the  one-electron reduced density matrix from the experimentally accessible Shannon entropy.

For spinless distinguishable electrons the entropy is sizeable at low energies,
indicating strong momentum entanglement, 
and drops to almost zero at energies of the order of 10 keV when the azimuthal degree of freedom is integrated out,
i.e. practically no entanglement and almost pure one-electron states.
If spin is taken into account, the entropy 
for electrons with antiparallel spins should be larger than in the  
parallel-spin 
case, since it embodies both  momentum and spin entanglement.
Surprisingly, this difference, as well as the deviation from the spin-less case,
is extremely small for the complete scattering state. 
Strong spin entanglement can however be obtained by post-selecting
states at scattering angle  $\pi/2$.
\end{abstract}
\end{frontmatter}


\section{Introduction}
If two or more particles interact with each other, the final many-particle state
is entangled, which means that it is no longer possible to attribute a complete set of 
properties to each of the particles (cf. e.g. \cite{Ghirardi-04a} and references therein).
This fundamental feature of quantum mechanics and its weird consequences were
first highlighted by the Einstein-Podolsky-Rosen (EPR) Gedankenexperiment \cite{EPR-35} 
and its discussion by Schr\"odinger \cite{Schroed-35}.
While originally of a more philosophical interest, entanglement is currently at the
heart of  important practical applications like quantum computing, quantum cryptography, 
teleportation (cf. e.g. review articles  \cite{Reid-09,Horodecs-09,Wineland-13,Adesso-17} 
and references therein).\\

In the case of  electron-electron collisions, the two-electron state after the interaction
can be entangled both with respect to spin and to momentum.
While a number of recent theoretical and experimental studies has focused on spin entanglement
\cite{Lamata-06b,Berezov-10,Petrov-14,Samar-15,Feder-15,Feder-17},
we are not aware of work on  momentum entanglement and its relation to
spin entanglement over a wide energy range. 
This is somewhat surprising, since many solid state and surface techniques
(such as e.g. electron energy loss spectroscopy (EELS), Auger  and  electron
microscopy) rely on information obtained by the Coulomb interaction of a
probe electron with the electronic system of the target.

Entanglement is closely related to the 'mixedness' of the state of the probe after interaction, and this in turn has to do with decoherence processes. 
 As an example we may take {\em elastic} scattering of electrons on a crystal which is considered as coherent whereas {\em inelastic} scattering such as ionisation is considered incoherent. As a consequence, the density operator of the probe electron after interaction is that of a pure state in the first case, and of a mixed state in the second case. 
As another confusing example of the complexity when coherent and incoherent processes occur simultaneously we mention plasmon scattering which is known to be delocalized over several nm, but plasmon filtered images show structure at atomic resolution. The latter observation has been explained quite early~\cite{Howie} as an artefact due to Bragg scattering of the probe electron before or after a plasmon loss event. 

More precisely --- apart of coincidence experiments~\cite{Kruit} --- it is the {\it reduced} density matrix of the probe that contains all information we can obtain on the status of the probed system. Reduced density matrices are closely related to entangled systems. Clearly, an interaction between formerly separated systems A (the probe, say) and B (the scatterer) creates entanglement. Sloppily,  to each state of A  belongs a state of B, so measuring a physical quantity on A tells us something about the state of B after the interaction. The strength or degree of entanglement is mirrored in the 'mixedness' of the reduced density matrix, which, in turn, determines the degree of coherence.


Based on the example of elastic scattering of electrons on a crystal mentioned above, it is tempting to believe that  elastic processes leave the  density matrices of the scattering partners pure. Interestingly, this is not the case for electron-electron scattering. 

In the following we discuss entanglement  in electron-electron scattering. After some basic definitions we use a simple model in order to obtain  the  mixedness  for spinless particles as a function of their kinetic energy, before  the spin degree of freedom is taken into consideration. We propose a scheme to estimate the  von Neumann entropy of the reduced one-electron density operator, based on the classical Shannon entropy in a momentum space representation. This provides insight not only   into electron-electron entanglement, but also into the information content of electron scattering experiments such as (e,2e)  or EELS~\cite{Schattschneider-UM2018}.

\section{Basic definitions} 
Density operators  are an extension of the concept of a wave function (pure state)  to an incoherent superposition of wave functions (mixed states):
$$
\hat \rho=\sum_i a_i a_i^* \ket{\psi_i} \bra {\psi_i}.
$$ 
When all except one of the coefficients vanish ($a_i=\delta_{i,0}$), the density operator reduces  to a single term, $\hat \rho= \ket {\psi_0} \bra {\psi_0}$. In this particular pure state the wave function describes the system completely. Trivially, the density matrix, (i.e. the representation of the density operator in an orthonormal basis) can be written as a product of the wave function with its complex conjugate, taken in that same basis.

Assume scattering between a plane wave electron and a target: The  probe is in a momentum eigenstate $\ket {\vec K} $ and the target state is denoted $\ket \Phi$. Before interaction the two subsystems  live in separate Hilbert spaces
$$
\ket{\psi_0}=\ket {\vec K} \otimes \ket \Phi.
$$
Long after the interaction of the two systems\footnote{That means longer than the interaction time, which for fast electrons to be considered later is of the order of $10^{-10}$ s.} 
  we may expand the total wave function into an orthonormal basis of the probe electron. Let us take a plane wave basis, postponing  the reason for this choice for the moment:
\begin{equation}
\ket{\psi}=\sum_i^N c_i \ket{\vec k_i} \otimes \ket{\phi_i} 
\label{eq:1}
\end{equation}
where we abbreviate the plane waves by  the scattering vector $k_i$ in the centre-of-momentum (CM) system. The $c_i$ are the excitation coefficients of  plane wave $\vec k_i$. For $N>1$ the two systems are said to be entangled, i.e. they cannot be separated because to any $\ket {\vec k_i}$ there belongs a $\ket{\phi_i}$. The wavefunction $\ket\psi$ cannot be written as a product of a single probe wave function with a single target wave function\cite{Landau1977}. The $\ket{\phi_i}$ are the states of the scatterer linked to the states of the scattered probe. The normalisation condition is $\sum_i |c_i|^2=1$. 
The density operator of the pure state $\ket{\psi}$ is
\begin{equation}
\hat \rho =\ket{\psi} \bra{\psi}=\sum_{i,j} c_i c_j^* \ket{\vec k_i} \ket{\phi_i} \bra{\vec k_j} \bra{\phi_j}.
\label{rho}
\end{equation}
For brevity  the $\otimes$ symbol is omitted here and in the following, taking the different notations for the two Hilbert spaces for granted. 

Quantitatively, the amount of information one can obtain on a system is encoded in the  density matrix $\hat \rho$ as the von Neumann entropy~\cite{Schlosshauer}
\begin{equation}
\mathcal S_N=-Tr[\hat \rho \log_2 \hat \rho].
\label{S}
\end{equation}
The density matrix of the probe-target system Eq.\ref{rho} is that of a pure state. When diagonalized, pure state matrices contain a single entry of probability one on the main diagonal, so its entropy Eq.~\ref{S} vanishes, i.e. we have complete knowledge of the system.  The quantum version of entropy --- Eq.~\ref{S} --- is very similar to the classical Shannon entropy\cite{Shannon1948}
\begin{equation}
	\mathcal S_S = -\sum_{i=1}^N p_i \, \log_2 p_i 
	\label{eq:Shannon}
\end{equation}
which is a measure of our ignorance of a (classical) stochastic variable $p$. In order to establish contact to the quantum case, the Shannon entropy relates always to a post selected basis with respect to which the probabilities are measured. If this basis happens to be an eigen basis of the density operator Eqs.~\ref{S} and \ref{eq:Shannon} coincide. 

$\mathcal S_S$ is closely related to the information content of a message. For instance, the Shannon entropy for fair coin tossing has $N=2$ with probabilities $p_{1,2}=0.5$ and is equal to one. That means there is one bit of information in each event (the result can be heads or tails). The relation to messaging is simply that to  transmit a tossing series of length $N$, $N$ bits are needed. Redundancies reduce the information content, so the same string of outcomes can be encoded with fewer bits. For instance, the title of this paper is a string of 56 characters, based on an alphabet of 18 symbols (space included). Encoding the message in bits means we need 5 bits to encode all symbols. For a completely stochastic arrangement of these 56 characters  that makes 5*56=280 bits. The  Shannon entropy of the text is 3.84~\cite{Kozlowski2018}. The redundancy in the text allows to represent the coded string with 216 bits.

In spite of their formal similarity there is a fundamental difference between the Shannon entropy $\mathcal S_S$  and the von Neumann entropy $\mathcal S_N$. In an experiment, one has access to $\mathcal S_S$. Only  in the eigen basis of the density operator $\mathcal S_N\equiv \mathcal S_S$. Unfortunately, this basis is difficult to approach experimentally, or not even known. Later on we discuss this important aspect for the actual case.

\section{Entropy of entangled systems}\label{sec:3}
After interaction, two formerly separated systems are entangled. They cannot  be described by two separate wave functions any more. The von Neumann entropy of such a bipartite system is still 0, but if only one of the two systems is observed, the ignorance of the second system introduces entropy that can be obtained from the {\it reduced} density operator of the first system after interaction. In the present case, the reduced density operator of the probe is, expanding the $\ket {\phi_i}$ into an orthonormal basis   and trace building 
\begin{equation}
\hat \rho_k=Tr_\phi[\hat \rho]=
\sum_{ij} c_i c_j^* \ket{ k_i} \bra{ k_j} \braket{\phi_j |\phi_i}
\label{rhok}
\end{equation}
as can be verified directly~\cite{Schlosshauer,Schattschneider-UM2018}.
When there is overlap of different states in the Hilbert space of the target, $\braket{\phi_j|\phi_i} \neq 0$, the probe electron shows interference properties in $k$ representation, i.e. it might still be coherent, at least to a certain  extent. If the overlap vanishes, i.e. if all target states are orthogonal to each other, the probe has decohered, it is in a mixed state\footnote{This is exactly the definition of pointer states in measurement theory: an orthonormal basis that is robust against interaction with the environment~\cite{Schlosshauer,Zurek1985}. In the present case, the environment against which the $\ket k$ basis is robust is the camera in the diffraction plane. This is the reason why we chose a plane wave basis.}. At the same time, the von Neumann entropy, Eq.~\ref{S} has increased.

The von Neumann entropy of a reduced density matrix measures the deviation of the subsystem from a pure state, or to put it differently, the degree of coherence. Since it is basis independent~\cite{Schlosshauer} we may write it in the {\it eigen} basis of $\rho_k$ in diagonal form
\begin{equation}
	\mathcal S_N=-\sum_i \lambda_i \log_2 \lambda_i .
	\label{eq:SS-SN}
\end{equation}

A pure state would have $\lambda_i=0$ except for one entry which by normalisation equals one, so $\mathcal S=0$. For an ensemble of $N$ pure states the maximum possible entropy is $\log_2 N$ when all states have the same probability. Similar to the classical entropy definition (here we use the logarithm to basis two, i.e. the Shannon entropy, but likewise the natural logarithm or the Gibbs entropy is common), it quantifies the amount of ignorance about the state of the probed system. In the present context, the extreme case of a pure probe state (no entanglement) is that of a completely coherent probe.  
Measuring a physical quantity of the probe does not tell us anything about the change of the state of the scatterer as there is no entanglement. For instance, after Bragg scattering the probe is in a coherent superposition of Bloch waves, and its entropy is zero. That means, repeating a diffraction experiment in an eigen (Bloch wave) basis always gives the same result, or we may say that there is no ignorance added by the interaction of the probe with the crystal.

\section{Spin-independent electron---electron scattering}

Let us begin with the simple case of scattering between two free plane wave electrons $ \ket{\vec K} \, , \, \ket{\Phi}$which are distinguishable from each other
and have no spin. The initial two-electron state is then just the simple product
$ \ket{\vec K} \, \otimes \, \ket{\Phi}$ and therefore not entangled. To get some touch of reality one might imagine an electron-electron collision in a storage ring or  in the electron microscope. In the center-of-momentum (CM) system the total wave function after interaction can be expanded as
\begin{equation}
\ket{\psi}=\int c_q  \ket{\vec K-\vec q} \ket{\Phi + \vec q} \, dq
\label{eq:2}
\end{equation}
with expansion coefficients $c_q$ that depend on the interaction. By definition of the system, $\Phi=-\vec K$ is a plane wave, and  $\vec q$ is the  scattering vector. The total momentum  is conserved in the interaction.
$\ket {-\vec K+\vec q}$ correspond to $\ket{\phi_i}$ in Eq.~\ref{eq:1}.
They are orthonormal, there is no mutual overlap, and hence the reduced density matrix  of the probe in a $\ket {\vec k}$ basis is, according to Eqs.~\ref{rhok},~\ref{eq:2}
\begin{equation}
\rho_k(\vec q, \vec q')=\expect {\vec K +\vec q |\rho_k | \vec K + \vec q'}= 
|c_q|^2 \delta^2(\vec q - \vec q')
\label{rhocont}
\end{equation}
where we have used the scattering vector $\vec q$ as the independent variable.
There are only diagonal elements, so we are in an eigen basis of the density operator. $\mathcal S_N$ can be measured in a plane wave basis.

The calculation seems to be straightforward, but we encounter a problem here.

\subsection{The problem of impact parameters}
The $c_q \propto q^{-2}$ in Eq.~\ref{eq:2} are the electron---electron scattering factors 
 which diverge for unscreened Coulomb interaction (e.g.~\cite{Dudarev1993})  for  $q \rightarrow 0$. The essential reason is the long-range behaviour of the Coulomb field. Nature has found a remedy by the simple fact that infinitely extended plane waves do not exist. The closest approximation to plane waves in the real world are wave packets with a finite lateral extension, (e.g. by some aperture, or by a focussing element, or by diffraction on periodic potentials). The divergence is removed because the finite lateral extension defines a maximum impact parameter $b$. Scattering angles in the azimuthally symmetric differential  scattering probability 
\begin{equation}
	\frac{d p}{d \Omega}=p(\theta)=|f(\theta)|^2
	\label{eq:dp}
\end{equation}
are then limited to~\cite{Demtroeder}
\begin{equation}
\theta > \epsilon=2 \, \arccot(2E \, b \frac{4 \pi \epsilon_0 }{e^2})
\label{thetamin}
\end{equation}
where $E$ is the total kinetic energy of the two electrons in the CM system. The corresponding minimum scattering vector is $q_{min}= K(E) \epsilon$ where $K(E)$ is the wave number of the electrons.

In the following, we use Gaussian wave packets\footnote{Note that in the stationary approach, the $z$ coordinate is eliminated; wave functions are defined in the plane perpendicular to the propagation direction of the electrons.} 
\begin{equation}
 \psi(\vec x) =e^{-x^2/4 \sigma^2}/\sqrt{2 \pi} \sigma
\label{eq:pure}
\end{equation}
for probe and target electron that collide head-on, an approach appropriate for  accelerators where such wave forms can approximately be realized. 
The extension of the wave packet is $L=2 \sigma$. 

The impact parameters are  distributed according to all possible  differences of positions in the two packets, weighted with the respective probabilities to find  the electrons there. This conditional probability is the convolution of the two packets:
$$
p(b) \propto e^{-b^2/4 \sigma^2}
$$
i.e. again a Gaussian. The standard deviation of b 
$$
\bar b=\sqrt{2} \sigma=L/\sqrt{2}
$$
serves here as an approximation for the limiting impact parameter between  the two Gaussian wave packets.  Inserting $\bar b$  into Eq.~\ref{thetamin}, the smallest scattering angle  is then defined   as a function of energy and wave packet extension.
The divergence at $\theta=0$ is now excluded, and the probability is  normalizable. 
The solution of the normalization problem comes at the price that plane waves are no eigen basis of the density operator any more. Instead we have extended wave packets that span the Hilbert space. In other words, $\mathcal S_N$ is not directly measurable in a $k$ basis. 

From the uncertainty relationship for Gaussian wave packets we have
\begin{equation}
	\sigma_k= \frac{1}{2 \sigma}=\frac{1}{L} .
	\label{eq:sigmak}
\end{equation}
The coherence length of wave packets in k-space   is $\sim 2 \sigma_k= 2/L$.
For large enough $L$, the packets are very narrow, filling only a narrow band along the main diagonal of the sparse matrix $\rho_k(\vec q,\vec q')$. In a way, the $k$ basis is {\it almost} an eigen basis. This gives a clue on how to obtain an approximate experimental value for $\mathcal S_N$ .

\section{Experimental estimate of $\mathcal S_N$}
\begin{figure}
	\centering
		\includegraphics[width= 0.9 \columnwidth]{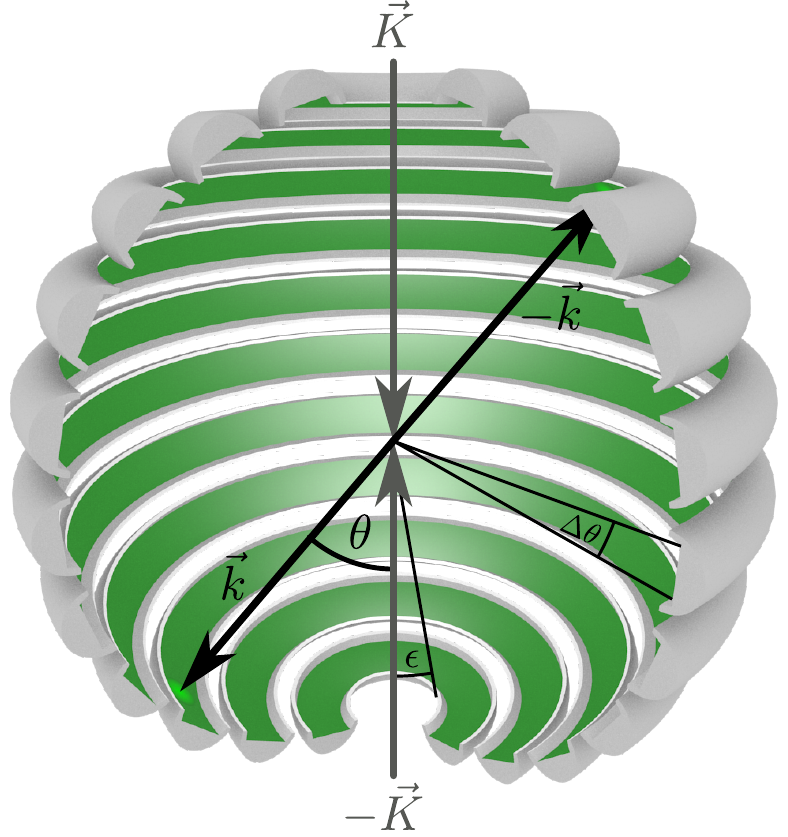}
	\caption{Scattering geometry with ring detectors. The relevant angles are indicated. The cattering vector is $\vec q=\vec k - \vec K$.}
	\label{fig:Fig1}
\end{figure}

For an eventual measurement, one may use a pixelated detector in the far field (i.e. the `almost diagonal' plane wave basis discussed above).  CMOS-based detectors with high resolution are standard in cameras for electron microscopy at energies of a few keV or higher, either as direct detection devices that offer single electron counting, or with a scintillator interface. Although they are conceived for forward scattering there is no reason why they could not be used for higher scattering angles. Special detector geometries can be realized by summing the signals of relevant pixels, for instance over the azimuthal angle.  Since the Coulomb scattering is azimuthally symmetric, it is reasonable to design ring detectors commencing at the minimum scattering angle $\epsilon$, extending up to $\theta=\pi$ with pixel width $\Delta \theta$ as sketched in Fig.\ref{fig:Fig1}. We call this detector setup a post selection. 
 The Shannon entropy  in this geometry depends on the choice of the number of pixels spanning the scattering range. 

The question is if there exists a pixel width that gives a Shannon entropy approximating the Neumann entropy of this post selection.
Intuitively, for the Gaussian wave packets the pixel width should be considered as the width of the Fourier transformed wave packet centered at $q_i$, $\Delta q_i=2/L$, or in terms of scattering angle $\Delta \theta_i=2/KL$. Defining diagonal coefficients\footnote{$f(q)$ relates to $f(\theta)$, Eq.~\ref{eq:dp} via $q=2 K \sin(\theta/2)$.} in polar coordinates $\vec q =(q,\phi)$
\[
c_i c_i^*:=2 \pi \int_{\Delta q_i} |f(q)|^2 \sin(q)\, dq 
\] 
with the now continuous scattering factor $f(\vec q) \propto q^{-2}$, 
the density matrix takes a discrete form
\begin{equation}
\rho_k \approx \begin{pmatrix}
c_1c_1^* & 0 &0\\
0 & c_2 c_2^*&0 \\
0 &0 & ...
\end{pmatrix} .
\label{rhodis}
\end{equation}
The pertinent consequence is that most of the off-diagonal elements will be zero or very small since we integrated over the coherence length of the packets. (This can also be understood from the uncertainty relationship which sets a lower limit for the useful pixel width  in $q$.)
The index labels the $i$-th pixel of the detector in the far field.
In order to see if the Shannon entropy 
\begin{equation}
\mathcal S_S=-\sum_i c_i c_i^* \log_2 (c_i c_i^*) 
\label{Entropy}
\end{equation}
is indeed a good approximation for the von Neumann entropy, we construct the reduced density matrix for the same post selection. 
The expression equivalent to Eq.\ref{rhocont}  for Gaussian wave packets  and trace building over the unobserved system yields
\begin{equation}
	\rho(\vec q, \vec q')=\int d^2q'' \rho_0(\vec q-\vec q'',\vec q'-\vec q'') |f({\vec q''})|^2
\end{equation}
where $\rho_0$ is the density matrix of the incident one-electron pure state with scattering vectors $\vec q, \vec q', \vec q''$ joining points on the Ewald sphere. We use the fact that $\rho_0$ is narrow and centred at $\vec q=0$ (it vanishes for arguments surpassing  a few $K \Delta \theta$), therefore we can replace the Ewald sphere  by its projection on the  plane tangent to  $\vec K$ (i.e. $\vec q=0$). The same holds approximately for $\vec q''$ because the main contribution to the integral comes from small $|\vec q''|$. 
With the Fourier transformed pure state Eq.~\ref{eq:pure} we obtain
\begin{eqnarray}
	\label{eq:rhoqq}
	\rho(\vec q, \vec q')&\approx & e^{- q^2/(4 \sigma_k^2)} e^{- q'^2/(4 \sigma_k^2)} \\
	&& \int d^2q'' e^{-(\vec q +\vec q') \vec q''/(2 \sigma_k^2)}e^{-q''^2/(2 \sigma_k^2)} |f({\vec q''})|^2 \nonumber.
\end{eqnarray}
Since the scattering is cylindrically symmetric, $f=f(|\vec q|)$,
\begin{eqnarray}
	\rho(\vec q, \vec q')&=&  e^{-{ q}^2/(4 \sigma_k^2)} e^{-{ q'}^2/(4 \sigma_k^2)}  \cr
	&&\int q'' \, dq''  e^{-q''^2/(2 \sigma_k^2)} \,|f({q''})|^2 \nonumber  \cr
&&	\int_0^{2\pi} \,d\phi \, e^{|\vec q+\vec q'|\, |\vec q''|\cos\phi/(2\sigma_k^2)} \nonumber 
\end{eqnarray}
with $\sigma_k$ given by Eq.~\ref{eq:sigmak}.
The azimuthal integral is the modified Bessel function of first kind $2\pi I_0(|\vec q+\vec q'|\, |\vec q''|/(2\sigma_k^2))$, so
\begin{multline}
	\rho(\vec q, \vec q') =  2 \pi e^{-{ q}^2/(4 \sigma_k^2)} e^{-{ q'}^2/(4 \sigma_k^2)}  \cr
	 \int q'' \, dq''   I_0(|\vec q+\vec q'|\, |\vec q''|/(2\sigma_k^2)  e^{-q''^2/(2 \sigma_k^2))} \,|f({q''})|^2.
	\label{2Dconv}
\end{multline}
In order to compare $\mathcal S_N$ with $\mathcal S_S$ we post select an infinitesimally narrow meridian.
Apart of a scaling factor  this selection is equivalent to the post selection which is used for $\mathcal S_S$ as sketched  in Fig.\ref{fig:Fig1}. On the meridian, the variable $|\vec q+\vec q'|$ is reduced to $|q+q'|$. The Neumann entropy  is obtained from Eq.~\ref{eq:SS-SN} after the numerical diagonalisation of the density matrix $\hat \rho_p$ of this polar post selection.
\begin{figure}[htb]
	\centering
		\includegraphics[width= \columnwidth]{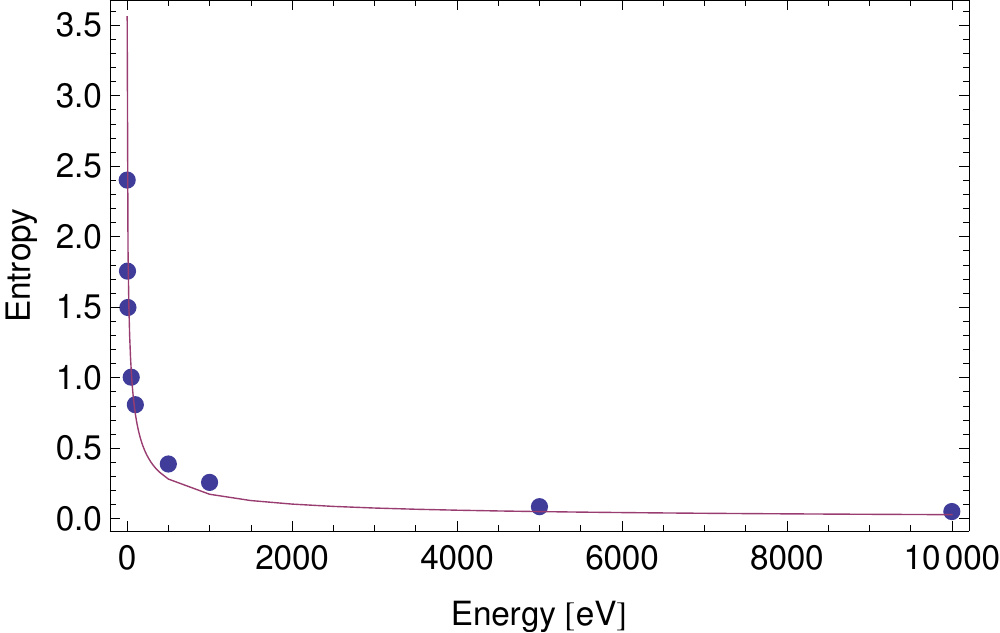} \\
		\includegraphics[width= \columnwidth]{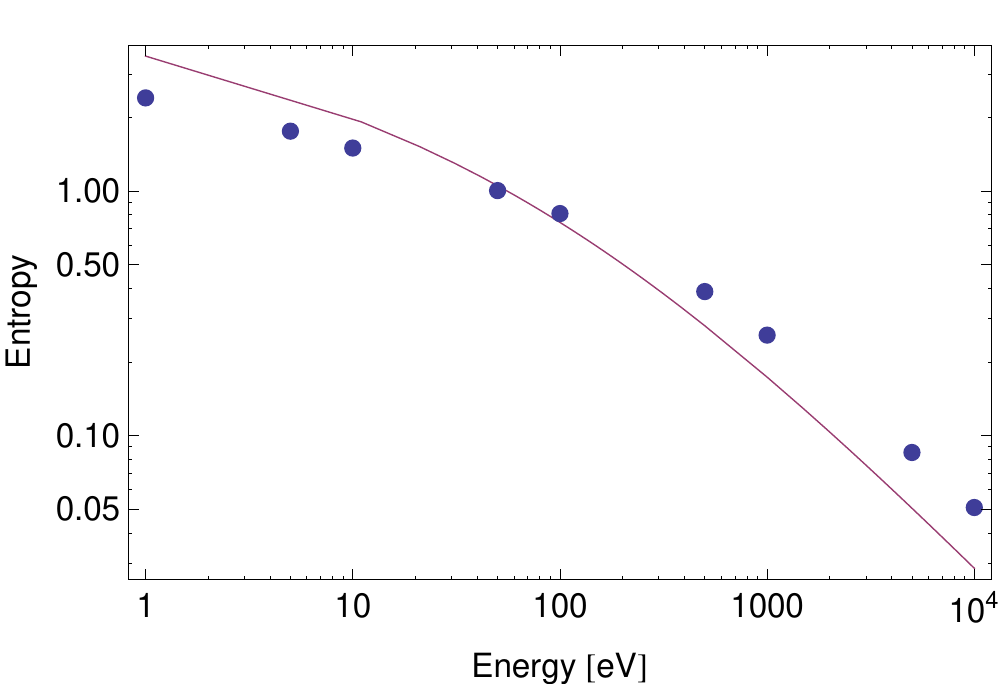}
	\caption{Shannon and von Neumann entropies for energies up to 10 keV for  ring detectors. a) linear plot; b) log-log-plot. The dots are Neumann entropies  obtained from diagonalising $\rho_p(\theta,\theta')$. Wave packet extension $L=100~nm$, detector width $\Delta \theta=2/(K L)$. }
	\label{fig:CombinedResultsSN-SS}
\end{figure}
Fig.~\ref{fig:CombinedResultsSN-SS} compares the Shannon entropy, Eq.~\ref{Entropy} with selected values of the von Neumann entropy for a range of energies and a Gaussian wave packet of extension $L=100$~nm. It should be mentioned that within the numerical accuracy the results do not depend on $L$. For the comparison, it is important to choose the same set of parameters for both $\mathcal S_N$ and $\mathcal S_S$. Both the Shannon and the von Neumann calculations suffer from an underestimation of coherence. The Shannon values are  too low for high energy because the coherence at small scattering angles is not taken into account properly, whereas the approximation Eq.~\ref{eq:rhoqq} becomes worse for low energies.   Nevertheless the overall agreement is good. It allows to use the simple Shannon discretisation for the later discussion.

\subsection{Numerical details}
For not too small kinetic energies in the collision ($\sim 100$~eV)  and not too narrow wave packets ($L \sim 10 \mu$m) the number of detectors is $> 10^5$.
This poses considerable numerical difficulties in the evaluation of Eq.~\ref{Entropy} that can be overcome with differential probabilities~\cite{Jaynes1957} derived in the Appendix. For $N$ ring detectors as sketched in Fig.~\ref{fig:Fig1} Eq.~\ref{eq:Jaynes} (see Appendix) gives
\begin{equation}
\mathcal S_{S,ring}=-2 \pi \int_\epsilon^{\pi-\epsilon} \bar p(\theta)  \log_2((\pi- \epsilon) \bar p(\theta)) \, d\theta + \log(N)
\label{eq:Sr}	
\end{equation}
where 
$$
\bar p=p(\theta) \sin(\theta)
$$
with the differential scattering probability $p(\theta)$, Eq.~\ref{eq:dp}.

 For completeness we also calculate the Shannon entropy for a pixelated detector with $M$  equidistant pixels covering the scattering sphere, each subtending a solid angle $\Delta \Omega=\Omega_0/M$
\begin{equation}
	\mathcal S_{S,sphere}=-2 \pi \int_\epsilon^{\pi-\epsilon} \bar p(\theta) \log_2(\Omega_0 \, p(\theta)) \, d\theta + \log(M)
	\label{eq:St}
\end{equation}
where $\Omega_0=4 \pi \cos\epsilon$
is the (incomplete) unit sphere.
\section{Spin-dependent electron-electron scattering}\label{spin-dep}

We now proceed from the case of two distinguishable particles without spin
to the scattering of two indistinguishable plane-wave electrons with momenta $\vec{K}$ and  $-\vec{K}$ 
(in the center-of-mass system)  and spin orientation  $\sigma_1 = \pm = up/down$ and $\sigma_2 = \pm = up/down$ 
with respect to some chosen axis. 
We first specify the relevant spin-dependent two-electron states and then address their 
entanglement and their entropies.

\subsection{Spin-dependent two-electron states}
The constituent one-electron states are written in the form $ |\vec{K},\sigma \rangle  := |\vec{K}\rangle ~|\sigma \rangle  $, 
where  $|\sigma \rangle$ is a Pauli spinor with  $\sigma_1 = \pm = up/down$.
The initial  two-electron state is then
\begin{equation}
\frac{1}{\sqrt{2}}  ( |\vec{K},\sigma_1\rangle  |-\vec{K},\sigma_2 \rangle - |-\vec{K},\sigma_2\rangle  |\vec{K},\sigma_1 \rangle),
\label{eq:spin-1}
\end{equation}
i.e. an  antisymmetrized product  of  two one-electron states, also referred to as a Slater determinant.

The final two-electron state $|ee\rangle$ after Coulomb interaction can be expressed in a plane-wave basis $|\vec{k}\rangle$ as
 \begin{eqnarray} 
|ee\rangle  &=&    \sum\limits_{k}^ N
       ( \frac{2 f_k}{\sqrt{2}}  ( |\vec{k},\sigma_1\rangle  |-\vec{k},\sigma_2 \rangle - |-\vec{k},\sigma_2\rangle  |\vec{k},\sigma_1 \rangle)   \cr
      &+& \frac{2 g_k}{\sqrt{2}}  ( |-\vec{k},\sigma_1\rangle  |\vec{k},\sigma_2 \rangle - |\vec{k},\sigma_2\rangle  |-\vec{k},\sigma_1 \rangle))   
			\label{eq:spin-2b}    
\end{eqnarray}
For each $\vec{k}$ the first and second terms represent direct and exchange scattering, respectively.  
The scattering amplitudes are (cf. e.g. \cite{Kessler-85})
\begin{eqnarray}
    f_{\vec{k}} &=&  1/ |\vec{k}-\vec{K}|^2 = 1/(4 K^2 \sin^2(\theta /2))  \nonumber
\\
g_{\vec{k}} &=&  1/ |-\vec{k}-\vec{K}|^2 = 1/(4 K^2 \cos^2(\theta /2)),
\label{eq:spin-3}
\end{eqnarray}
where $\theta$ is the scattering angle in the center-of-mass system.
In Eq \ref{eq:spin-2b}  the symmetry relation
\begin{equation}
 f_{-\vec{k}} = g_{\vec{k}}
\label{eq:spin-3b}
\end{equation}
has been taken into account by multiplying the scattering amplitudes by the factor 2 and by summing only over half of the
sphere of radius K (energy shell). More precisely,
the $\vec{k}$ summation comprises momenta $\vec{k}$ with   $\epsilon \leq  \theta  < \pi/2$ and $0 \leq \phi  < 2\pi$,
and  $\vec{k}$ with $\theta = \pi/2$ and $0 \leq \phi  < \pi$. $N$ is the number of the momenta $\vec{k}$ thus selected for
a given discretization,  i.e. only half of the number of momenta compared to the case of distinguishable particles.

Since scattering cross sections and entropies of the final two-electron state  $|ee\rangle$ turn out to be different for
parallel and for anti\-parallel spins, these two cases have to be considered separately.
For parallel spins, i.e.  $\sigma:=\sigma_1= \sigma_2=\pm$, Eq.\ref{eq:spin-2b} reduces to
\begin{equation}
         |ee\rangle_{par}  = \sum\limits_{k}^N
       2 (f_k-g_k)  \frac{1}{\sqrt{2}} ( |\vec{k},\sigma \rangle  |-\vec{k},\sigma  \rangle - |-\vec{k},\sigma \rangle  |\vec{k},\sigma  \rangle).
\label{eq:spin-4}
 \end{equation}
For antiparallel spins, i.e. $\sigma:=\sigma_1=\pm$ and $\bar\sigma:=-\sigma= \sigma_2=\mp$, 
the final two-electron state has the form
\begin{eqnarray}
     |ee\rangle_{ap}  &=&  \sum\limits_{k}^N    ( 
       \frac{2f_k}{\sqrt{2}}  ( |\vec{k},\sigma \rangle |-\vec{k},\bar\sigma \rangle - |-\vec{k},\bar\sigma \rangle |\vec{k},\sigma \rangle) \cr		
    &-&  \frac{2g_k}{\sqrt{2}}  ( |\vec{k},\bar\sigma \rangle |-\vec{k},\sigma \rangle - |-\vec{k},\sigma \rangle |\vec{k},\bar\sigma \rangle)
     ).
\label{eq:spin-5}
\end{eqnarray}
%
%
\subsection{Entropy and Entanglement}
The entropy of  the above  two-electron states (Eqs \ref{eq:spin-1},\ref{eq:spin-4},\ref{eq:spin-5})
can be calculated in the same way as in the spinless case.
The corresponding two-electron density matrices are traced out with respect to
momentum and spin coordinate of one electron to obtain the one-electron reduced
density matrices $\rho$.
Since this calculation is straightforward but rather lengthy, it may suffice to show
the results.

In the parallel-spin case we obtain from  Eq \ref{eq:spin-4}
the one-electron reduced density matrix

\begin{equation}
      {\rho}_{par} =  
    \left(   \begin{array} {cccccc}
                  c_1c_1^* & 0 & 0 &0 &0 &0 \\
                   0 & c_1 c_1^*  &  0 &0  &0 &0 \\
                  0 & 0 & c_2c_2^* &0  &0 &0 \\
                  0 & 0 & 0 & c_2 c_2^* &0 &0  \\
                  0 & 0 & 0 & 0 &  .. & 0\\
                  0 & 0 & 0 & 0 & 0 & .. \\
   \end{array}  \right), 
\label{eq:rho_par}
\end{equation}
where
\begin{equation}
    c_k : =  (f_k-g_k) / \sqrt{ 2  \sum\limits_{k'}^N |f_{k'} -g_{k'}|^2}.
\label{eq:spin-9}
\end{equation}
The Shannon entropy is then
\begin{equation}
       \mathcal S_{par} =  - \sum\limits_{k=1}^N  2 \, |c_k|^2 \log(|c_k|^2). 
\label{eq:spin-9b}
\end{equation}
In contrast to the case of two distinguishable particles without spin (cf. Eq 11), 
$\rho_{par}$ contains two diagonal elements for each $\vec{k}$.
The reason for this is the antisymmetry of the two-particle wave function.
This is most clearly seen by considering a state with just one momentum  $\vec{k}$,
like the initial state Eq.~\ref{eq:spin-4} with parallel spins ($\sigma_1$  = $\sigma_2$).
The reduced density matrix is then (2x2) with elements 1/2 on the diagonal.
and the entropy is $\mathcal S$=1. The same is obtained if one removes the spins from Eq 19. 
In contrast to distinguishable particles without spin with one $\vec{k}$ , 
the indistinguishable ones with parallel spins and even without spins have
entropy $\mathcal S$=1 instead of $\mathcal S$=0.  
Naively one might therefore say that the two indistiguishable electrons are -- 
even without any interaction -- entangled, whereas distinguishable ones are not.
As has been discussed extensively in the literature
(\cite{Schliemann-01,Ghirardi-04a,Lamata-06b} and references therein), this is however
not adequate from a physical point of view.
Instead of $\mathcal S_{par}$,  the modified entropy 
\begin{equation}
\tilde{\mathcal S}_{par} :=  \mathcal S_{par} -1
\label{eq:spin-9c}
\end{equation}
is a suitable measure for the genuine
entanglement \cite{Ghirardi-04a} brought about by interaction.
Information-theoretically speaking, the ignorance embodied in $\mathcal S$ is split up into a Pauli ignorance with
entropy 1 and a Ghirardi ignorance with entropy $\tilde{\mathcal S}_{par}$,
the latter being the physically  relevant one.
Accordingly, the above two-electron state with just one momentum $\vec{k}$, which has  $\tilde{\mathcal S}_{par}=0$,
is not genuinely entangled, whereas states, which cannot be expressed
as a single antisymmetrized product of two one-electron states, are genuinely entangled.

In the antiparallel-spin case we obtain from  Eq \ref{eq:spin-5}
the one-electron reduced density matrix
\begin{equation}
      {\rho}_{ap} =  
    \left(   \begin{array} {cccccc}
                   \tilde{f}_1  \tilde{f}_1^* & 0 & 0 &0 &0 &0 \\
                   0 & \tilde{g}_1  \tilde{g}_1^*  &  0 &0  &0 &0 \\
                  0 & 0 & \tilde{g}_1  \tilde{g}_1^* &0  &0 &0 \\
                  0 & 0 & 0 & \tilde{f}_1  \tilde{f}_1^* &0 &0  \\
                  0 & 0 & 0 & 0 &  .. & 0\\
                  0 & 0 & 0 & 0 & 0 & .. \\
   \end{array}  \right), 
\label{eq:rho_par}
\end{equation}
where 
\begin{equation}
   \tilde{f}_{\vec{k}}  =  f_{\vec{k}}   / \sqrt{2 \sum\limits_{k'}( |f_{\vec{k'}}|^2  + |g_{\vec{k'}}|^2)}
\label{eq:spin-5b} 
\end{equation}
and
\begin{equation}
    \tilde{g}_{\vec{k}}  = g_{\vec{k}}   / \sqrt{2 \sum\limits_{k'}( |f_{\vec{k'}}|^2  + |g_{\vec{k'}}|^2)}
\label{eq:spin-5c} 
\end{equation}
The entropy of  ${\rho}_{ap}$ is obtained as
\begin{equation}
        \mathcal S_{ap} =  \sum\limits_{k}^N (
         2 |\tilde{f_k}|^2\log_2  (|\tilde{f_k}|^2) +2 |\tilde{g_k}|^2 \log_2 ( |\tilde{g_k}|^2) ).
\label{eq:spin-9d1}
\end{equation}
Like in the above case of parallel spins, the modified entropy
\begin{equation}
     \tilde{\mathcal S}_{ap} : =   \mathcal S_{ap}  -1
\label{eq:spin-9d}
\end{equation}
then characterizes the genuine entanglement. While for parallel spins there
is only momentum entanglement, the antiparallel case generally exhibits
both momentum and spin entanglement. 
The latter is most prominent in
an antiparallel-spin state of the form  of Eq. \ref{eq:spin-5} with only a single momentum $\vec{k}$,
which is chosen such that the scattering angle $\theta$ is 90$^o$.
Since according to Eq.\ref{eq:spin-3} one then has $f_{\vec{k}}=g_{\vec{k}}$, 
this state is the paradigmatic Bell singlet state~\cite{Bell-64} with $\tilde{\mathcal S}_{ap}$=1.

Comparing  the entropies for parallel spins with those for antiparallel spins,  one can expect the latter to be generally larger
than the former, because they represent spin entanglement in addition to momentum entanglement.

A connection with Coulomb scattering of  two distinguishable particles (with equal masses) can be established in the above 
anti-parallel spin case. If one employs spin filters such that for momenta $\vec{k}$  and $-\vec{k}$ one detects
only electrons with spin $\sigma$ and $-\sigma$, respectively, the two electrons can be regarded as distinguishable 
by their spin label and the  final two-electron state Eq.\ref{eq:spin-5} reduces to the direct part  without antisymmetrization
and with the $\vec{k}$ sum covering the entire energy shell (except for the $\epsilon$ cone around  $\theta=0$):
\begin{equation}
     |ee\rangle  =    \sum\limits_{k}^{2N}  f_{\vec{k}}  |\vec{k}\rangle   |-\vec{k}\rangle .
\label{eq:spin-12}
\end{equation}
The entropy is then obtained in terms of the normalized scattering amplitudes 
$\tilde{f}_{\vec{k}} = f_{\vec{k}}  / \sqrt{  \sum\limits_{k}  |f_{\vec{k}}|^2}$ 
as
\begin{equation}
   \mathcal S=  -  \sum\limits_{k}^{2N}  |\tilde{f_k}|^2\log_2 |\tilde{f_k}|^2  .
\label{eq:spin-13}
\end{equation}

 {\section{ Results and Discussion}
\subsection{Spinless case}
Fig.~\ref{fig:2} shows results for the spinless case, Eq.~\ref{Entropy} as a function of the total kinetic energy from 1 to 100 eV (sum of both particles' energy in the CM system; in the lab frame, the probe energy is twice as high). For the Gaussian wave packet  with L=50~nm,  the entropy decreases rapidly with the kinetic energy of the particles. At 100~eV  the two electrons show almost no entanglement; the probe is still highly coherent because the largest part of the scattering amplitude is in the forward direction, and that means that the original incident wave packet remains almost unchanged after scattering. The entropy is  low because the probability that the target remains in its original state is  high when the   forward scattering probability is  high.
\begin{figure}[htbp]
	\centering
		\includegraphics[width= \columnwidth]{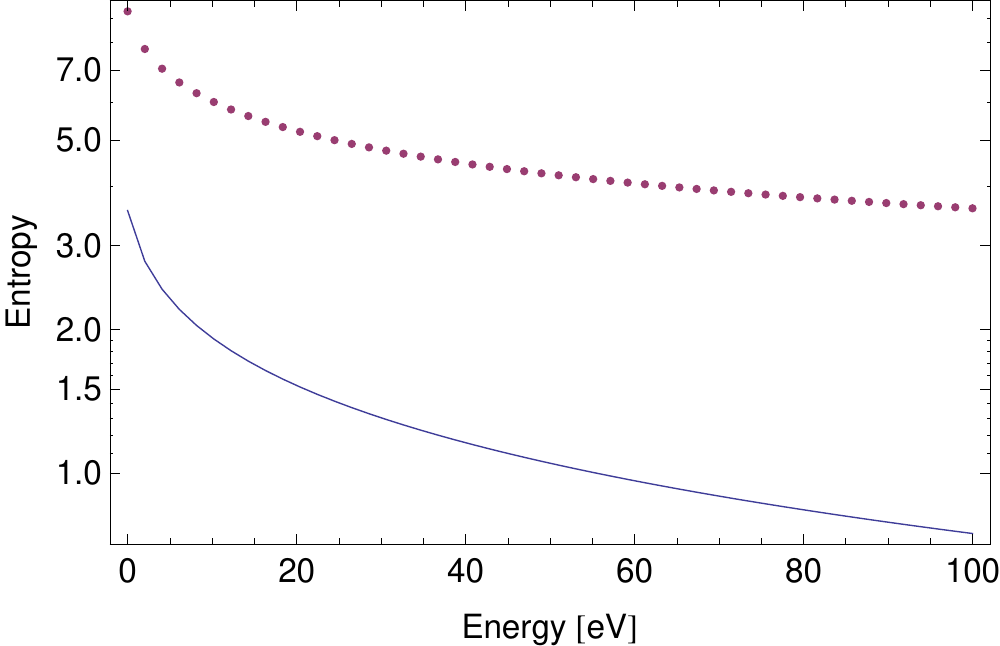}
	\caption{Shannon entropy for el-el collisions in the CM system. Full line: $\mathcal S_{S,ring}$ (ring detectors collecting signal over the entire azimuth); dotted line: $\mathcal S_{S,sphere}$ (upper bound for equidistant distribution of detector pixels over the  unit sphere, Eq.~\ref{eq:St}). 
	}
	\label{fig:2}
\end{figure}
For 100~eV, $\Delta \theta=0.039$~mrad, that makes 80700 detectors covering the whole scattering angle, and $8.3 \cdot 10^9$ pixels covering the sphere.
Quantitatively, the information encoded in a scattering experiment at energy 1 eV is 3.5 bits, or to put it differently, less than 4 bits are needed on the average to describe a particular experimental outcome in terms of scattering angle. The same experiment performed at 100 eV contains only  $\sim 0.7$ bits of information because 8 out of 10 detection events will be at the smallest discrete  scattering angle defined by the maximal impact parameter, and 2 eventw  at the second and higher ones. We note that at 50 keV, equivalent to 100 keV in the laboratory frame, a typical kinetic energy of a TEM-probe, the entropy drops to $\sim 7.0 \, \cdot  10^{-3}$. The information contained in a scattering experiment is now highly redundant - only one out of 2000 scattered electrons goes to the second largest angle, and not a single electron lands at higher angles. There is practically no entanglement. The figure shows also the entropy $\mathcal S_{St}$ Eq.~\ref{eq:St} for equidistant detectors covering the whole unit sphere. It must be cautioned that these values represent just  upper bounds for the Neumann entropy which is much lower, especially for higher energy. The reason is that the equidistant detector arrangement poses at least three detectors at the smallest scattering angle, which all have the same probability of detection, so the Shannon entropy is always $> 1.5$  whereas the reduced density matrix at high energy is almost pure.

\subsection{Spin-dependent case}


Next we present entropy results for spin-dependent electron-electron scattering.
From the equations in Sec.~\ref{spin-dep}, one can ---  without numerical calculations --- already draw 
some general conclusions. 

In the case of anti-parallel spins the modified Shannon entropy $\tilde{\mathcal S}_{ap}$ (Eq \ref{eq:spin-9d})
reflects the combination of spin and momentum entanglement. 
  To shed some light on their relative weights 
it is useful to postselect the half-ring of width $\Delta  \theta$ centred at $\theta = \pi /2$ 
subdivided into $N_r = \pi / \Delta  \theta$ azimuthal detector elements (pixels).
Thus there are $N_r$ different momenta $\vec{k}$ in the azimuthal interval  [0.. $\pi$]. 
Out of this set, we further select $N$  momenta  (with  1 $\leq N \leq N_r$).
 Since for $\theta=\pi/2$   we have  $f_k = g_ k  $  (cf. Eq.\ref{eq:spin-3}), 
the modified Shannon  entropy for antiparallel spins is readily obtained from Eq.\ref{eq:spin-9d} 
as 
\begin{equation} 
\tilde{\mathcal S}_{ap} =  \log(2N)  = 1 + \log N   
\label{eq:spin-antipar}
\end{equation} 
The first (second) term  characterizes the spin (momentum) entanglement.
Thus for $N=1$ there is only spin entanglement with $\tilde{\mathcal S}_{ap} = 1$. 
With increasing $N$, spin entanglement is seen to be increasingly masked by momentum 
entanglement.

For parallel spins, there is only momentum entanglement.
For $\theta = \pi /2$, the  effective signal in a single detector $\Delta \theta$ is
$$
\int_{\pi/2-\Delta \theta/2}^{\pi/2+\Delta \theta/2}|f_k - g_k|^2 d\theta.
$$
This is extremely small, according to Eq.\ref{eq:spin-3}. 
Nevertheless, the entropy can be derived from the fact that the probability distribution for  $N$ detectors on the equator is uniform. 
One thus obtains
\begin{equation} 
 \tilde{\mathcal S}_{par} = \log N .   
\label{eq:spin-par}
\end{equation} 
Consequently, the entropy difference between the antiparallel and the parallel spin case is 1, which is
the signature of spin entanglement.
Taking into account all the $N_r$ detectors on the ring, this difference is still 1, but momentum entanglement
produces a much larger entropy.
For example   for $\Delta  \theta$ = 1 mrad we have $N_r$=3140 and Eq.\ref{eq:spin-par}  yields  $\tilde{\mathcal S}_{ap}$=11.6 . 
For $\Delta \theta=0.17$~mrad, which we used in the numerical calculations,
$N_r$=18050 and $\tilde{\mathcal S}_{ap}$=14.1 .

Analytical results can also be obtained for individual  $\theta_i$ pixels post-selected along a
meridian. {\it i. e.} for fixed azimuth, $\phi$=0, say). Since the post selected momentum is fixed there is no momentum entanglement, only spin entanglement for antiparallel spins.
Going from $\theta= \pi /2$  towards $\epsilon$, the scattering factor $f(\theta_i)$  increasingly
dominates over $g(\theta_i)$  (cf. Eq.\ref{eq:spin-3}). Consequently $\tilde{S}$ decreases
from its maximal value 1  towards 0.
This case was studied in detail  by \cite{Lamata-06b} and \cite{Samar-15} for free electrons
and by \cite{Feder-15,Feder-17}  for  electron-induced pair emission from solid surfaces.

The dominance of $f$ over $g$ at small scattering angles together with its steep increase
has an important consequence for the overall entropies (obtained from all pixels on the
sphere). 
Firstly, the differences between the three entropies (no spin, parallel spins and antiparallel spins)
are very small.
Secondly, the three entropies tend towards zero in the high-energy limit, since the minimal
scattering angle tends towards zero.

Numerical calculations for the spin-dependent cases quantitatively corroborate these
expectations.
The entropy for parallel-spin  electrons (Eq. \ref{eq:spin-9b}) is almost identical with  its
counterpart obtained for two distinguishable particles ("spinless electrons" or electron and
positron) (Eq. \ref{eq:spin-13}).
The difference is $<10^{-5}$ in the  energy range 1 eV - 10 keV. 
The entropy for antiparallel spins (Eq.\ref{eq:spin-9d}) differs only by $\sim 10^{-10}$ or 
less from the parallel case (Eq. \ref{eq:spin-9b}) in the same energy range.
Spin entanglement, which exists only for antiparallel spins, is thus completely
masked by momentum entanglement, which exists in all three cases.
Given such minute differences, plots of the spin-dependent entropies are visually
identical with the plots of the spinless entropy shown in Figs.~\ref{fig:CombinedResultsSN-SS} and \ref{fig:2}.

To understand in more detail the reason for this  surprising result, 
we calculated the spin-dependent entropies for post-selected angular ranges $\theta \in [\pi/2-\theta_r, \, \pi/2]$
as functions of  $\theta_r$ going from 0 to $\pi/2-\epsilon $.
Results obtained for an energy of 5 eV are shown in Fig. ~\ref{fig:EntropyNew-1b}.
The angular spread of one detector ring is $\Delta \theta=0.174 $~mrad, giving 9000 hypothetical  
channels when the whole scattering range, $\theta \in [\epsilon,\pi/2]$ is covered. 

In Fig. 4a the entropy for antiparallel spins is seen to clearly exceed the parallel-spin one,
with the difference entropy  $\Delta \mathcal S$ (dashed line) of the two exhibiting a wide
plateau of $\Delta \mathcal S\approx 1.6$ up to about 1~rad. 
The classical result, Eq.~\ref{Entropy} is shown as a dotted line in between the two.
It is closer to the antiparallel case at larger angles and closer to the parallel case at  
smaller angles. 

Fig. 4b shows a zoom into post-selections up to only 100 pixels away from $\pi $/2.
For the first pixel close to $\pi/2$, for which there is no momentum entanglement,
we find $\mathcal  S_S=0$ for the spinless and the parallel-spin case, and $\mathcal  S_S=1$ for the antiparallel one 
due to spin entanglement.
Post-selecting the first two pixels, the corresponding entropy values are 1, 0.54, and 2. 
If both pixels carried equal weights, one would expect a momentum entanglement entropy
of 1,  i.e. all three entropies should exceed their one-pixel counterparts by 1.  While this is
actually so for the spinless and the antiparallel case, it surprises at first sight that the 
parallel-spin entropy is only 0.54.  The reason for this is that the coefficients $(f_k-g_k)^2$ 
rise rather steeply from the value 0 at $\pi$/2 so that the coefficient of the second pixel  is
significantly larger than the one of the first pixel. 
In contrast, the coefficients $c_k^2$ in the spinless case and  $f_k^2$ and $g_k^2$ in the
antiparallel case are almost the same in the first two pixels.

For large angular range, all three curves in Fig. 4a converge to the same limiting value of $\mathcal S_S \approx 2.4$, 
i.e. there is no difference between the spin-endowed and the classical spinless case.
This is due to the fact that near the minimum scattering angle $\epsilon$
the exchange coefficients $g_k$ almost vanish, whereas the direct coefficients $f_k$ and the "spinless"
coefficients $c_k$ are very large, concentrating most of the weight in a few pixels near  $\epsilon$.
The convergence of the three curves is demonstrated in more detail in Fig. 4c by the difference 
entropies in post-selections of pixels near the minimum scattering angle.



%
\section{Conclusion}

	We have studied momentum and spin entanglement in el-el Coulomb interaction. 
To this end, we have established a simple discretisation method that allows a reasonable  
estimate of the von Neumann entropy of the reduced one-electron density matrix from the measured 
Shannon entropy in the far field of the interaction region. 
The vast number of pixels resulting from this particular  discretisation demands 
the adaption of the continuous entropy concept of Jaynes to the actual situation. 

For 'spin-less' and for parallel-spin electrons, there is only momentum entanglement.
In the antiparallel-spin case there is also spin entanglement, which is however
severely masked by momentum entanglement except 
under favourable post selection of states. We extend the findings of 
Lamata {\em et al.} \cite{Lamata-06b} showing 
how and under which conditions spin-independent Coulomb interaction can cause spin-dependent entanglement.
	
As functions of energy, upper bounds for the Shannon entropies of  $\sim 9.3 $  at 1 eV, decreasing   to   $\sim 1.8$  at  10 keV have been calculated. It must however be warned that for high energies, the approximation $\mathcal  S_S \approx S_N$ breaks down. 

For a post selection along a meridian of the unit sphere the values are $\sim 3.5 $  at 1 eV  to   $\sim 0.03$  at  10 keV.
This means that  post selection along a meridian at very high energies suppresses momentum entanglement; the outgoing one-electron state is almost pure. This observation is useful in EELS where model calculations are almost exclusively based on pure states.
Conversely, in order to study and to employ momentum entanglement, one has to go to very low energies.
\begin{figure}[htb]
	\centering
		\includegraphics[width=\columnwidth]{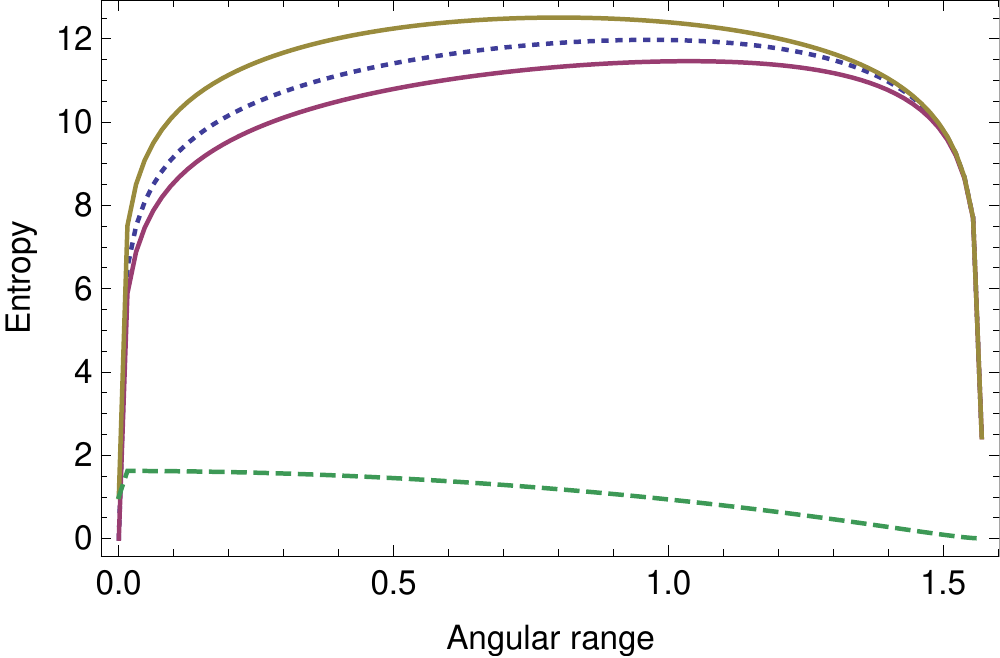}
	\vskip 0.4 cm
		\includegraphics[width= \columnwidth]{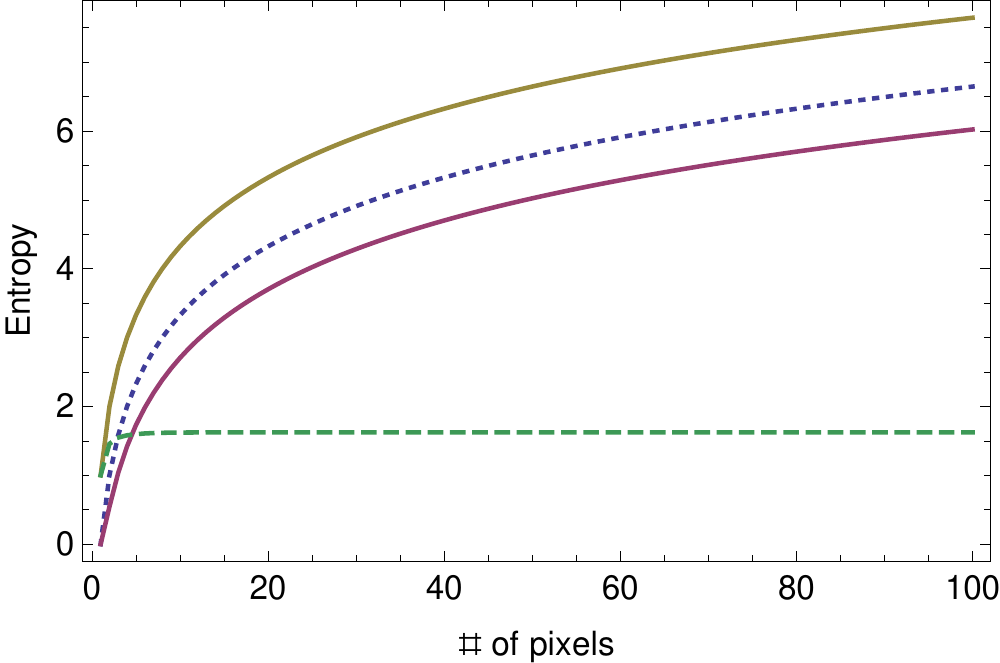}
		\vskip 0.4 cm
		\includegraphics[width= \columnwidth]{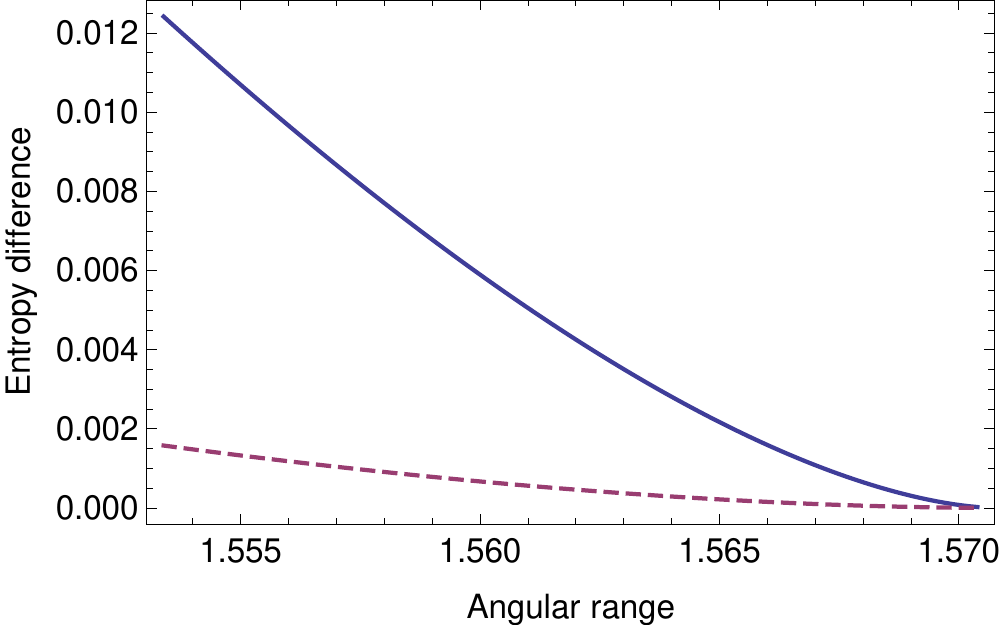}
		
	\caption{a) Spin-dependent entropies as a function of post-selected angular scattering range $[\pi/2-\theta_r, \pi/2]$. 
	Parameters:  $E0=5$~eV, Gaussian model ($L=100$~nm), resulting in $\Delta \theta=0.174$~mrad. The whole angular range corresponds to more than 9000 pixels (discrete detectors). Full lines: anti-parallel (higher values)  and parallel spins (lower values). Dotted: spinless case. Dashed: difference antiparallel minus parallel.
	b) Zoom-in to a given number of pixels close to $\theta=\pi/2$.  c) Zoom-in to the rightmost part of a). Full line: difference antiparallel minus parallel, Dashed: difference antiparallel minus spinless.  }
	\label{fig:EntropyNew-1b}	
\end{figure}

\section*{Acknowledgements:}
  P. S. acknowledges the financial support of the Austrian Science Fund under grant number P29687-N36.

\section*{Appendix}

Assume a pixelated detector covering the unit sphere with each pixel subtending a solid angle $\Delta \Omega=\Delta \theta^2$ . The differential scattering probability Eq.~\ref{eq:dp}
\[
	\frac{d p}{d \Omega}=p(\theta)=|f(\theta)|^2
\]
vanishes for $\theta < \epsilon$, as given in Eq.~\ref{thetamin}. We also exclude scattering angles $\theta>\pi-\epsilon$ for symmetry reasons that become important in the case of indistinguishable particles.
The probability that one pixel "fires" is
$$
p_i = \int_{\Delta \Omega} |f(\theta)|^2 \, d\Omega:= p(\theta_i) \Delta \Omega,
$$
where $p$ is  normalized on the unit sphere as
$$
\int_0^{2 \pi} \int_{\epsilon}^{\pi-\epsilon}{p(\theta) d \Omega}=2 \pi \int_{\epsilon}^{\pi-\epsilon}{p(\theta) \sin(\theta) \, d\theta}=1.
$$
In spherical coordinates, there are 
$$
m_i= \frac{2 \pi \sin(\theta_i)\Delta \theta}{\Delta \Omega}
$$
detectors at $\theta_i$. The Shannon entropy is
\begin{eqnarray}
\mathcal S_S&=&-\sum_i^N p_i \, m_i \log(p_i)= \\
&=&-\sum_i^N p(\theta_i) \Delta \Omega \frac{2 \pi \sin(\theta_i) \Delta \theta}{\Delta \Omega} \log(p(\theta_i) \Delta \Omega) \nonumber 
\label{S2D}
\end{eqnarray}
where  $N=(\pi-2\epsilon)/\Delta \theta$.
We rewrite the logarithm  as
$$
\log(p(\theta_i) \Delta \Omega)=\log(p(\theta_i) M \Delta \Omega) -\log(M) .
$$
Here, $M=\Omega_0/\Delta \Omega$ is the number of detectors on the (incomplete) unit sphere
$$
\Omega_0=4 \pi-2 \cdot 2\pi(1-\cos\epsilon)=4 \pi \cos\epsilon .
$$
We obtain
\begin{eqnarray}
\mathcal S_S&=&-\sum_i^N 2 \pi \, p(\theta_i)  \sin(\theta_i) \log(p(\theta_i) M\Delta \Omega) \, \Delta \theta + \nonumber \\
&+&\sum_i^N 2 \pi \, p(\theta_i)  \sin(\theta_i) \log( M) \, \Delta \theta.
\end{eqnarray}
The second term is $\log(M)$. We replace the first term with the limit $\Delta \Omega \to 0$, as shown by Jaynes~\cite{Jaynes1957}, and since $\lim_{\Delta \Omega \to 0}(M \Delta \Omega)=\Omega_0$ 
 the result is
\begin{equation}
	\mathcal S_S=-2 \pi \int_\epsilon^{\pi-\epsilon} p(\theta) \sin(\theta) \log(\Omega_0 p(\theta)) \, d\theta + \log(M).
	\label{eq:Jaynes}
\end{equation}
The case of ring detectors, Eq.~\ref{Entropy} can be worked out in complete analogy, resulting in Eq.\ref{eq:Sr}, with the number of pixels $M$ replaced by the number of rings $N$. 
 As an aside we note that in the particular case of an isotropic probability distribution, $f(\theta)=const.$  so $\Omega_0 \, p(\theta)=1$, and the first term vanishes. Since uniform probability distributions maximize the entropy, we may interpret the continuous (Jaynes) entropy as the reduction of entropy for non-uniform probability distributions. 


\end{document}